\pdfoutput=1  
\documentclass[11pt]{article}
\usepackage[margin=1in]{geometry}
\usepackage{setspace}
\usepackage{needspace}
\usepackage{graphicx,booktabs,amsmath,url}

\usepackage[font={small,stretch=1.0}]{caption}
\usepackage{xcolor}
\usepackage[authoryear,round]{natbib}
\usepackage[colorlinks=true,linkcolor=blue,citecolor=blue,urlcolor=blue]{hyperref}

\usepackage{etoolbox}
\makeatletter
\patchcmd{\HyOrg@maketitle}
  {\hb@xt@1.8em{\hss\@textsuperscript{\normalfont\@thefnmark}}}
  {\@textsuperscript{\normalfont\@thefnmark}\,}
  {}{\patchcmd{\maketitle}
       {\hb@xt@1.8em{\hss\@textsuperscript{\normalfont\@thefnmark}}}
       {\@textsuperscript{\normalfont\@thefnmark}\,}
       {}{\typeout{WARNING: thanks-footnote indent patch failed}}}
\makeatother

\newcommand{\dropcap}[1]{#1}
\newcommand{\SIname}{Appendix}

\newcommand{\singlefigwidth}{0.55\linewidth}
\newcommand{\kwsep}{, }
\newcommand{\availability}{The analysis pipeline and the pre-commitment record are publicly archived at OSF, \url{https://osf.io/gqw9b} (DOI: \url{https://doi.org/10.17605/OSF.IO/GQW9B}).}
\newcommand{\precommitrecord}{The core-analysis predictions (P1--P3) were committed at \texttt{53e5637} and the staleness-test addendum (P5) at \texttt{8355945}, in each case before any estimation script for those analyses had been written. The record is archived at OSF (\url{https://osf.io/gqw9b}): the predictions document exactly as of each cited commit, the git ordering evidence, and the full analysis pipeline; the archive is public, with DOI \url{https://doi.org/10.17605/OSF.IO/GQW9B}.}
\newcommand{\predictionsdoc}{Full statements are in \texttt{docs/ideation/predictions.md} within the repository snapshot archived at OSF, at the commits cited in the text.}
\newcommand{\repostatement}{The full pipeline (taxonomy construction, curation with quoted evidence, panel builds with hard-failing integrity assertions, estimation, and figure scripts that assert their inputs against the recorded results) is archived, with exact package versions pinned, at OSF (\url{https://osf.io/gqw9b}; DOI: \url{https://doi.org/10.17605/OSF.IO/GQW9B}), together with the pre-commitment record referenced in section~\ref{si:predictions}.}

\newcommand{\papertitle}{Stranded Credentials: \\ Keeping Online Reputation Systems Informative in the AI Era}
\newcommand{\paperkeywords}{credentials\kwsep signaling\kwsep generative AI\kwsep evaluation design\kwsep online platforms}

\title{\papertitle}
\author{Song Yao\thanks{Olin Business School, Washington University in St.\ Louis. E-mail: \texttt{songyao@wustl.edu}. The author holds concurrent appointments as Professor of Marketing at the Olin Business School, Washington University in St.\ Louis, and as an Amazon Scholar. This paper describes work performed at Washington University in St.\ Louis and is not associated with Amazon.}}
\date{\today}

\begin{document}
\maketitle

\begin{abstract}
\singlespacing

Platforms summarize providers' past achievements into credentials that buyers use to judge quality. Generative AI can now produce much of the work those achievements certify, raising fears that those quality signals are worthless. We audit how well such credentials stay informative in Kaggle's 2010--2026 archive, where medals are won on predictions scored against withheld answers and two evaluation formats ran side by side. Across 444,698 participations, a medal's power to predict performance sits almost entirely in its first year, in both formats and eras. Fresh medals kept most of their value through the AI transition. About half of the collapse in the informativeness of one format's medals is institutional: the platform had been phasing out that format before AI, and its medal stock aged out on schedule. Old medals look more informative only in isolation. The platform's official lifetime-tier display discards up to a sixth of the medals' predictive power. A recency-weighted index fit before the AI era explains AI-era performance about 13\% better than the tiers and selects entrants who perform better on average, though the tiers still identify extreme top performers better. Displaying a recent-performance summary alongside the lifetime tiers would recover the discarded information for buyers.

\end{abstract}

\section{Introduction}

\dropcap{P}latforms that summarize providers' past achievements into credentials must keep those credentials informative for buyers as AI changes how work is produced. Freelance marketplaces award badges for highly rated deliverables, review platforms rank reviewers by the quality of their past reviews, and question-and-answer sites rank members by their accepted answers. Generative AI can now draft, at minimal cost, the analysis, the code, and the essay that many such credentials were designed to certify \citep{hui2024employment}. When the artifact behind a credential becomes cheap, the credential may certify something different from what it did, or nothing at all. When AI tools are readily accessible, the cost of producing these artifacts no longer separates the skilled from the unskilled, and unverified skill signals already show the damage: written proposals on freelance platforms stopped predicting effort once AI could write them \citep{galdin2025talk}, AI-polished pitches became harder to screen \citep{cowgill2026cheaptalk}, and unproctored assessment scores inflated where generative AI was accessible \citep{eduassess2026validity, ited2026mcq}. A common conclusion is that skill credentials may become worthless for analytics work \citep{ashton2026vibe}. Whether credentials backed by verified performance remain informative in the AI era, and how a platform should summarize such credentials, are open questions. Answering both requires scored outcomes before and after the AI transition. The second question is older than AI. The transition is a major shock under which that question can be tested out of sample.

We audit such a credential system end to end. Kaggle, the largest data science competition platform, has recorded 18.7 million competition entries by hundreds of thousands of participants since 2010 \citep{boenisch2025kaggle}, and awards medals and lifetime tiers (Expert, Master, Grandmaster) that members display as professional credentials. Every medal is won on predictions scored against withheld answers, so the credentials certify verified performance rather than artifacts. Two features make the platform an unusually clean setting. First, two evaluation formats ran side by side: \emph{upload-competitions}, which score submitted predictions and cannot tell whether the process behind them was ever validated to be generalizable, and \emph{code-competitions}, which run entrants' programs on hidden data to produce out-of-sample predictions, so a process that does not generalize is caught. The platform does not prohibit AI assistance in either format; the two formats differ in whether skipping the validation has consequences \citep{tadelis2026analytics}. Second, the platform retired the upload format for reasons that predate generative AI, which lets us watch one credential stop being issued while its stock stayed on display. Throughout, we measure one construct, the \emph{informativeness} of a credential: how well it predicts subsequent hidden-test performance. Informativeness is a property of the measurement system, not of the people measured. The estimates are associational. Whether AI changes human skill is unidentifiable here by construction, because entrants are never observed without AI access, and we make no such claim.

Four findings follow. None matches the fear that credentials are now worthless, and each informs how a platform should summarize a participant's verified history into the credentials it displays. First, medals are short-lived signals: in every era and format, nearly all of a medal's predictive power sits in its first year. Second, the platform's lifetime tiers discard up to a sixth of the information in the medals. A recency-weighted index fit on pre-AI outcomes alone explains AI-era performance better than the tiers. The index also selects entrants who perform better on average, while the tiers remain better at identifying the extreme top performers. Third, about half of the collapse of the upload-earned credential stock is institutional stranding: once the platform had phased out that format, the stock aged out under the pre-existing decay pattern. Fourth, old medals look more informative in the AI era only when read in isolation, because a stale medal proxies for its holder's other signals, and the person-level changes we had predicted did not happen: within person, an AI-like working style predicts performance similarly in both formats, and execution failures rose with what competitions became, not who the entrants are. Several of these findings emerged by rejecting predictions we had pre-committed to a version-controlled record before estimation (\SIname~A1).

This study contributes to research on reputation and certification systems on platforms. Since \citet{spence1973job}, credentials have been understood as signals that survive only while they stay correlated with what they certify. The platform literature studies how feedback and certification thresholds shape market outcomes \citep{tadelis2016reputation, dranove2010quality, hui2025certificates} and how ratings inflate or get manipulated \citep{filippas2022inflation, he2022fake}. It rarely observes the predictive content of a credential directly, and it almost never watches a credential stop being issued while its holders' later performance keeps being scored. Both are observed here. The results complement, rather than contradict, the evidence that AI degrades unverifiable signals \citep{galdin2025talk, cowgill2026cheaptalk}. Kaggle medals were never cheap talk, and they stayed informative through the same technology shock. The damage that did occur is a matter of design: how a platform weights aging credentials, whether it retires the format that issued those credentials, and how it displays what remains.

\section{Setting, data, and methods}\label{sec:methods}

\textbf{Data and sample.} Meta Kaggle, the platform's public archive (snapshot of July 29, 2026), records 18,668,885 competition submissions from 2010 to 2026. The analysis panel covers 444,698 person-competition participations by 197,561 users in 214 medal-eligible upload- and code-competitions (the latter introduced in 2019; other formats are excluded, \SIname~A2) with deadlines from 2018 to 2025, excluding the 2022Q4 transition quarter when ChatGPT was released; the 2010--2017 years enter only as credential histories, and 2026, incomplete at the snapshot, is excluded. Entrants compete alone or in teams (solo in 73\% of participations). When a team competes, every member is credited with the team's result. This choice does not drive the findings: the informativeness estimates are the same when the sample is restricted to solo entrants and when a team's result is credited only to the member who submitted it (\SIname~A4).

\textbf{Outcome and credentials.} The outcome is one minus a team's final percentile on the hidden-test leaderboard, so higher values mean better performance. The predictors are the participant's medal counts prior to the competition, grouped into bands by how long ago each medal was won and which format awarded it, and entered as $\log(1+m)$. Official tiers are reconstructed from the platform's published deterministic thresholds and validated against observed tiers (97.7\% accuracy; \SIname~A5). The platform also publishes a points ranking that decays with a half-life of about a year (\SIname~A2); we reconstruct it as of each competition's start and benchmark it in section~3.2. The tier is the more visible credential: it marks a member's profile and username across the site, while the ranking appears only on the member's profile page and a leaderboard.

\textbf{Eras and estimation.} The pre-AI era comprises competition deadlines through 2022Q3, the quarter before ChatGPT's release, and the AI era deadlines from 2023 onward. Participation-level least squares with competition fixed effects confine every comparison to entrants facing the same task; standard errors are two-way clustered by user and competition. Each credential variable is interacted with an AI-era indicator, so its slope can differ between the two eras, and the interaction coefficient measures the change in that credential's informativeness. \availability{}

\section{Results}

\subsection{Credentials are short-lived signals}

\begin{figure*}
\centering
\includegraphics[width=\linewidth]{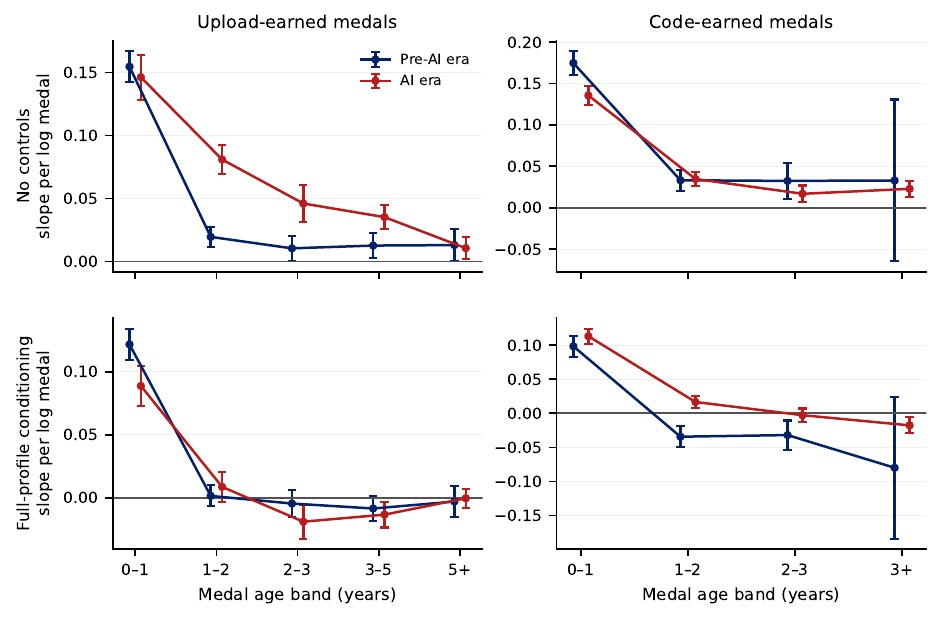}
\caption{Medal informativeness concentrates in the first year in both eras, with and without the full credential profile. Slopes of hidden-test performance on log medal counts per medal age band, from participation-level regressions with competition fixed effects, estimated separately for pre-AI-era outcomes (competition deadlines 2018--2022Q3) and AI-era outcomes (2023 onward); whiskers are 95\% confidence intervals, two-way clustered by user and competition. Top row: no further controls. Bottom row: conditioning on the participant's full credential profile (the other format's medal stock and prior participation count). Columns: upload-earned and code-earned medals. Code-earned medals older than three years are merged into one point (3+), since code medals were first awarded in 2019. The AI-era rise of old upload-earned medals in the top row is the profile proxying analyzed in section~\ref{sec:revaluation}.}
\label{fig:decay}
\end{figure*}

Fig.~\ref{fig:decay} shows how the informativeness of a medal varies with its age, estimated separately for the two eras. Each band's slope is the weight that medals of that age carry in predicting performance: a large slope means one such medal moves the prediction substantially, and a slope near zero means the medal contributes nothing to the prediction. The decay in informativeness is steep, and it is not an artifact of any particular specification. Unconditionally (Fig.~\ref{fig:decay}, top row), a \emph{fresh} medal, less than a year old, carries a pre-AI-era slope of 0.15--0.18 per log medal (one fresh medal versus none predicts performance 10--12 percentile points better), while medals older than one year retain slopes of only 0.01--0.03. This concentration of informativeness in the first year is unchanged when the full credential profile (the participant's other medals and prior participation) is taken into account (Fig.~\ref{fig:decay}, bottom row). The concentration is a statement about information content beyond a slope comparison: when predicting performance with all medal age bands of both formats, the two under-one-year bands alone deliver 99\% (pre-AI era) and 95\% (AI era) of the within-competition variance explained by the full set of bands (\SIname~A5). Nearly all of the predictive power of a Kaggle medal lives in its first year. The pattern predates AI, holds in both formats, and reappears when the profiles are re-estimated on AI-era outcomes (the AI-era series in Fig.~\ref{fig:decay}). We do not attribute the decay to any single mechanism: the holder's skill or effort may fade, or the data science tools and tasks may change, and none of these mechanisms is visible to a reader of the record. Everything below builds on the informativeness decay, regardless of its cause: a credential \emph{stock} is only as informative as the recent medals replenishing it.

\subsection{The platform's summary discards information}

\begin{figure*}
\centering
\includegraphics[width=\linewidth]{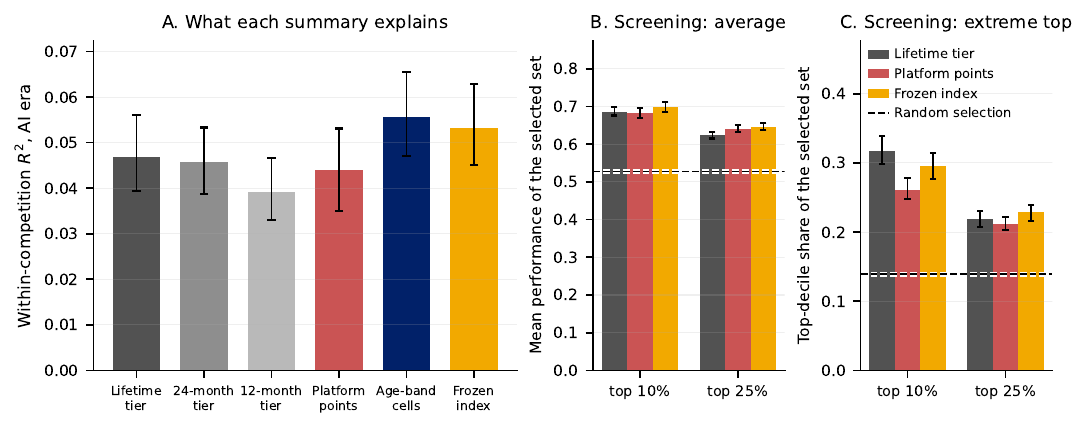}
\caption{The official summary discards recoverable information, and recovering it improves a broad screen. (\emph{A}) Within-competition $R^2$ of hidden-test performance on AI-era outcomes for the official lifetime tier, the same rule restricted to medals from the last 24 or 12 months, the platform's decaying competition points reconstructed as of each competition's start (log scale), the age-band-by-format medal counts, and the recency-weighted index with weights fit on pre-AI outcomes only and then frozen (the frozen index); whiskers are competition-cluster bootstrap 95\% intervals ($B=200$). Intervals for the frozen index's differences from the tier and from the points exclude zero (\SIname~A5). The index $R^2$ treats the frozen index as a single regressor: its nine age-band-by-format weights stay at their pre-AI values, and only the slope on the index is estimated on AI-era outcomes. (\emph{B}) Mean performance and (\emph{C}) share of top-decile finishers of the entrants selected by each rule at the 10\% and 25\% cutoffs per competition, averaged across competitions, with entrants tied at a cutoff sharing the remaining slots equally; dashed lines are the values among all entrants (random selection).}
\label{fig:screening}
\end{figure*}

Kaggle's official tiers are deterministic functions of lifetime medal counts, computed from thresholds the platform publishes.\footnote{\texttt{kaggle.com/progression}, accessed on September 12, 2026.} Those thresholds allow us to reconstruct any user's credential tier at any date (97.7\% accuracy against observed current tiers, with a 1.3\% false-positive rate among untiered users; \SIname~A5). As a predictor, the tier is a rule never fit to outcome data: a medal counts the same regardless of its age or format, and the cumulative totals collapse into four classes (untiered, Expert, Master, Grandmaster). The decay pattern of Fig.~\ref{fig:decay} implies that the age-blind totals must waste information. The open questions are how much information the totals waste, whether recovering it requires anything the platform lacked before the AI transition, and what the recovery is worth in a screening decision.

Against the official tier we construct a recency-weighted index built from the same medal events, with an estimated weight for each age-band-by-format cell. In sample, where the weights are estimated on the same era's outcomes they are evaluated on, replacing the tier with those age-band-by-format cells raises within-competition predictive power by 15\% (pre-AI era) and 19\% (AI era); equivalently, the official aggregation discards 13\% (pre-AI era) and 16\% (AI era) of the information in those cells. Out of sample, the index, with weights estimated entirely on the platform's pre-AI-era outcomes and then frozen, outperforms the tier on AI-era outcomes (within-competition $R^2$ 0.053 versus 0.047, about 13\% higher; competition-cluster bootstrap 95\% confidence interval for the difference, 0.004--0.009; Fig.~\ref{fig:screening}A). Accordingly, improving on the platform's own credential rule requires no information the platform lacked before the AI transition.

The platform's own decaying points ranking is the natural incumbent for a recent-performance summary and a benchmark for our recency-weighted index. We therefore reconstruct competition points as of each competition's start from the published formula and validate the reconstruction against members' current points (93\% within 5\%; \SIname~A2 and~A5). On AI-era outcomes the frozen index explains more than log points (within-competition $R^2$ 0.053 against 0.044; paired difference 0.009, 95\% interval 0.004--0.013; Fig.~\ref{fig:screening}A), and points explain no more than the lifetime tier (0.044 against 0.047, an interval that includes zero). The index therefore recovers information that neither displayed credential carries on its own.

The age-band-by-format cells give the regression more parameters than the tier (nine cells against four classes), which invites one objection: the cells might win the in-sample comparison through flexibility alone. First, flexibility cannot explain the out-of-era gain, since frozen weights cannot overfit the evaluation sample. Second, the cells contain information the tier lacks: when the tier and the cells enter the same regression, the cells still add three to four times more predictive power on top of the tier than the tier adds on top of the cells (within-$R^2$ increments of 0.013 and 0.012 by the cells against 0.004 and 0.003 by the tier across the two eras; \SIname~A5).

Two further exercises, not pre-committed (\SIname~A1), separate what the index recovers from how a platform could use it. First, imposing recency through the tier's own thresholds does not recover the gain. Recomputing the official rule on trailing windows, so that only medals from the last 24 or 12 months count toward a tier, explains no more than the lifetime rule (within-competition $R^2$ 0.046 or 0.039 against 0.047 in the AI era; the 12-month rule is substantially worse), but a short window under lifetime thresholds also empties the upper classes, so the exercise cannot separate recency from coarseness (\SIname~A5). Second, a screening exercise translates the gain into a decision. Within each AI-era competition we select the top 10\% or 25\% of entrants by the lifetime tier and by the frozen index constructed using pre-AI outcomes, with identical set sizes, and compare the selected entrants' subsequent performance (Fig.~\ref{fig:screening}B and C). Entrants selected by the index perform better on average at both cutoffs (mean performance 0.699 versus 0.686 at 10\%, and 0.645 versus 0.624 at 25\%; the bootstrap 95\% intervals for both differences exclude zero), and at the 25\% cutoff more of them finish in their competition's top decile (0.228 versus 0.219). At the 10\% cutoff, however, the lifetime tier selects more top-decile finishers (0.318 versus 0.295): a cut so narrow is filled largely by the two highest classes, whose members reach the top decile most often, whereas the index spreads the same cut over more classes. Lifetime honors remain the better instrument for identifying the extreme top performers, while the index improves a broader screen. The platform's points, added as a third selector, trail the index on both outcomes at both cutoffs (mean performance 0.684 against 0.699 at 10\%; top-decile share 0.262 against 0.295) and select fewer top-decile finishers than the lifetime tier (0.262 against 0.318 at 10\%). All three selectors far exceed random selection, whose top-decile share is 0.139.

\subsection{Stranding explains about half of the upload-earned stock's collapse}

Between 2016 and 2024, the platform's medal-eligible competition lineup flipped from 26 upload- and 0 code-competitions per year to 3 and 24: the platform effectively phased out the upload format in favor of code execution, amid scrutiny of public-leaderboard overfitting that predates AI \citep{roelofs2019overfitting}. We call the upload-earned medal stock \emph{stranded}, as a coal power plant is stranded when the grid switches to renewable energy: devalued by the phase-out of the evaluation format awarding the medal, not by any measured change in the medal holder's performance. 

\begin{figure*}
\centering
\includegraphics[width=\linewidth]{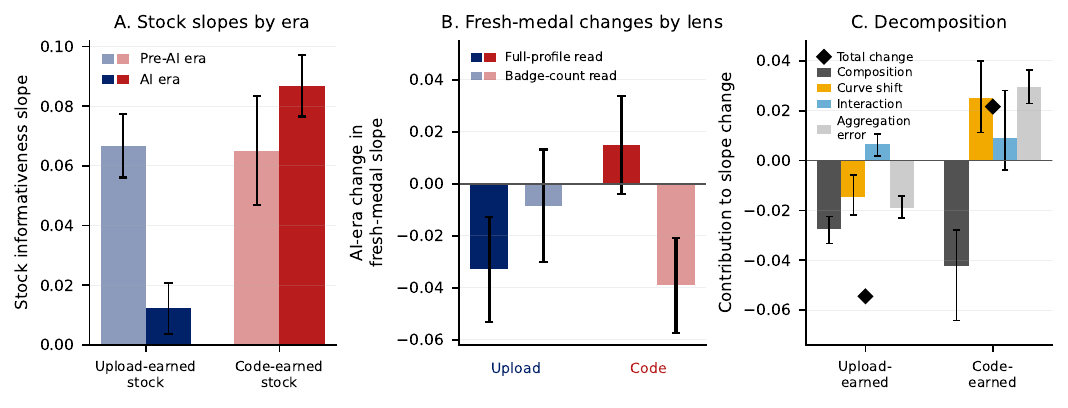}
\caption{The upload stock's decline traces more to aging than to revaluation. (\emph{A}) Informativeness slopes of the upload-earned and code-earned medal stocks (per log medal, conditional on the other stock and on prior participation count, with competition fixed effects) for pre-AI-era versus AI-era outcomes; whiskers are 95\% confidence intervals. (\emph{B}) AI-era changes in fresh-medal slopes by format under the full-profile read (dark) and the badge-count read (light). (\emph{C}) Decomposition of each stock's slope change into a composition component, a curve component, their interaction, and the aggregation error that closes the identity; diamonds mark each stock's total change; whiskers are competition-cluster bootstrap 95\% intervals per component.}
\label{fig:stranding}
\end{figure*}

Fig.~\ref{fig:stranding}A shows what the phase-out of the upload format did to the two medal stocks as credentials. Before the AI transition, upload-earned and code-earned medal stocks were similarly informative (slopes of 0.067 and 0.065 per log medal, each conditional on the other stock and on prior participation count). For AI-era outcomes, the slope of the upload-earned stock fell by 0.055 (s.e.\ 0.007, $p<0.001$) to 0.012, a loss of 82\%, while the code-earned stock rose by 0.022 (s.e.\ 0.010, $p=0.04$) to 0.087. These are full-sample changes, the relevant estimand for a buyer, who faces the whole AI-era pool including newcomers. Among users who competed in both eras, where continued participation holds by construction, the upload-earned stock's decline is $-0.019$ (s.e.\ 0.006, $p=0.001$) and the code-earned stock's rise $+0.014$ (s.e.\ 0.009, $p=0.13$; \SIname~A4). If AI had compressed the field (narrowing performance gaps) or reshuffled it (re-ranking who does well), every signal's slope would move together; instead, the two stocks move in opposite directions within the same competitions, a pattern that a uniform compression or reshuffling of the field would not produce. The same divergence across evaluation formats appears in community competitions, which award no medals, so the pattern is not specific to the medal-eligible sample (\SIname~A4). Despite the devaluation, the stranded medals are still displayed at face value on the platform's profiles and lifetime tiers. 

The decomposition of slope changes in Fig.~\ref{fig:stranding}C, an Oaxaca--Blinder-style split of each stock's slope into its age-band shares and band slopes (\citealp{oaxaca1973male, blinder1973wage}; \SIname~A4), shows that only a quarter of the upload-medal slope decline is the medals themselves being revalued; about half is the decay pattern of Fig.~\ref{fig:decay} manifested in an aging stock. Of the $-0.055$ change, $-0.028$ comes from the stock's age mix shifting as it aged in place while replenishment dwindled (the composition component), $-0.014$ from the band slopes themselves moving at the pre-AI mix (the curve shift), $+0.007$ from the two moving together (the interaction), and the remaining $-0.019$ is band-level aggregation error (sums may differ by rounding). The aggregation error ($-0.019$) is, however, larger than the curve shift ($-0.014$). The decomposition thus leaves more of the decline unexplained than it attributes to the medals being revalued. Aging alone, the composition component, accounts for about half of the decline (51\%; competition-cluster bootstrap 95\% interval 42--61\%), and it exceeds the curve shift in every resample. The interaction and the aggregation error are properties of the decomposition rather than of either mechanism, and we attribute them to neither. The same decomposition does not describe the code-earned stock, whose composition, curve, and interaction terms roughly offset while the aggregation error exceeds the total change (Fig.~\ref{fig:stranding}C): its rise is not because of aging. It reflects the re-weighting of the surviving signal analyzed in the next section, interacting with a rapidly changing age mix.

How a credential is read matters throughout this market, so we distinguish two reads used in everything that follows: a \emph{badge-count read} looks at one credential in isolation (the unconditional slope), while a \emph{full-profile read} weighs each credential given everything else on the record, including the other format's medals and prior participation. The full-profile read is the one that matters when comparing candidates, because it asks what each signal adds beyond the rest of the record. Genuine AI-era declines in fresh-medal informativeness exist but are moderate and depend on the read (Fig.~\ref{fig:stranding}B): under the full-profile read, fresh upload medals lost about a quarter of their slope ($-0.033$ on a pre-AI base of 0.122 from Fig.~\ref{fig:decay}, bottom left, $p=0.002$) while fresh code medals did not decline ($+0.015$, $p=0.12$); under the badge-count read the pattern reverses ($-0.009$, $p=0.44$, against $-0.039$, $p<0.001$). The reversal is proxying at work, analyzed in the next section.

\subsection{Old medals look more informative only when read alone}\label{sec:revaluation}

\begin{figure*}
\centering
\includegraphics[width=\linewidth]{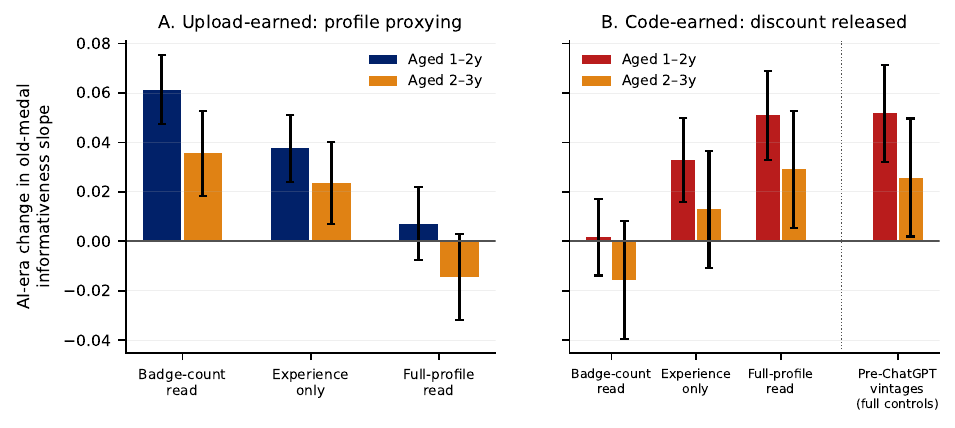}
\caption{Old-medal informativeness changes with the read, in opposite directions by format. Bars are AI-era minus pre-AI changes in slope for medals aged 1--2 and 2--3 years, with 95\% confidence intervals, as controls move from the badge-count read to the full-profile read; the middle group conditions on prior participation only. (\emph{A}) Upload-earned medals: the badge-count gains vanish under the full-profile read. (\emph{B}) Code-earned medals: gains appear only under conditioning and undo pre-AI negative weights. The rightmost group re-estimates the full-profile change for medals earned before ChatGPT, with post-ChatGPT code medals controlled. Not pre-committed (\SIname~A1).}
\label{fig:consolidation}
\end{figure*}

Did the medals of the surviving format become more durable, retaining value with age and defying the decay pattern? The two reads give opposite answers. Fig.~\ref{fig:consolidation} shows that every apparent era change in the value of \emph{old} medals is a property of the read, not of the medal, with the direction reversed between the two formats. This decomposition was not pre-committed; we added it after questioning our own earlier results (\SIname~A1). Under the badge-count read, old upload-earned medals look \emph{more} informative in the AI era (slope changes of $+0.061$ and $+0.036$ for medals aged 1--2 and 2--3 years); the gain in informativeness shrinks gradually as controls are added and vanishes under the full-profile read (Fig.~\ref{fig:consolidation}A).

The natural suspect is survivorship: credentialed participants were far more likely to keep competing into the AI era (7.8\% of unmedaled pre-AI-era participants appear in AI-era outcomes, against 56.6\% of those with six or more medals). The balanced panel, however, speaks against this explanation: among users with at least one participation in each era, where continuation holds by construction, the gains are unchanged ($+0.061$ and $+0.037$, both $p<0.001$; \SIname~A4), although the panel holds neither exposure nor participation frequency fixed. What inflates stale medals under the badge-count read is instead the rest of the holder's record: a medal read in isolation proxies for its holder's other signals, above all fresh code medals, and full-profile conditioning absorbs exactly that proxying. For code-earned medals the opposite pattern appears (Fig.~\ref{fig:consolidation}B). The reason is that the two credentials reflect the same underlying ability and hence compete to explain the same outcome variation: each carries only partial weight in a full-profile read, and when one signal stops predicting, the weight assigned to the other rises even if nothing about it changed. 

The estimated weights show this mechanism at work. Before the AI transition, old code medals carried significantly \emph{negative} full-profile weights ($-0.034$ and $-0.032$ for the two age bands, the below-zero points in Fig.~\ref{fig:decay}, bottom right): a discount for overlapping with the upload stock and prior participation, the stronger signals about the same participant (for the arithmetic, \SIname~A4). As an analogy, imagine two recommendation letters from referees who always agree: the second letter adds little beyond the first. Once the more discerning referee's letter is in hand, the other letter matters only where it deviates, and any praise that the discerning referee does not echo can even count against the candidate. In the AI era the old code medals' weights rise by $+0.051$ ($p<0.001$) and $+0.029$ ($p=0.017$), to $+0.017$ and $-0.003$ (sums may differ by rounding): the discount vanishes, rather than the old medal becoming a strong positive signal in its own right. Medals earned \emph{before} ChatGPT show the same re-weighting ($+0.052$, $p<0.001$, and $+0.026$, $p=0.035$, controlling post-ChatGPT code medals; Fig.~\ref{fig:consolidation}B, rightmost group), so a changed pool of medal earners cannot explain the gains. In short, there is no evidence that any medal type became intrinsically more durable or more informative; what changed is the portfolio of signals around each medal.

Two person-level changes that we had predicted and pre-committed did not appear (\SIname~A6). Within person, an AI-like working style, measured from submission telemetry with the index of \citet{yao2026scaffold}, predicts performance no differently across the two formats ($+0.006$ per index unit, s.e.\ 0.005), and the rise in failed runs among entrants arriving after AI tools tracks the competitions rather than the entrants: competition fixed effects reduce the AI-era entrants' excess failure from $+0.056$ to $-0.003$. These are nulls for a working style, not for AI use, which is never observed.

\section{Implications and limitations}

What do hard-earned credentials still measure in the AI era, when AI tools are readily accessible? Roughly what they always measured, for roughly as long: a medal's predictive power still concentrates in its first year, the credentials explain about a fifth less within-competition variation than before (the tier's $R^2$ fell from 0.059 to 0.047), and the AI era changed how the signals were issued and read more than their informativeness. The AI era therefore enters as a test of the credentials' informativeness rather than as a mechanism that changes it. The three design lessons below hold in both eras. What the AI transition adds is evidence that the lessons hold when the work behind the credentials changes, and that the advantage of a recency-weighted summary over the official tier survives the transition. Generative AI made polished submissions cheap to produce, but it did not make a top-rank credential cheap to achieve. Kaggle medals were never cheap talk: in either format, a medal is won on predictions verified against withheld answers. Their trajectory therefore shows what verification does not protect against: even verified signals decay fast, lose informational value when the type of evaluation that awarded them is retired, and re-weight with the signals around them.

\needspace{4\baselineskip}
The three platform design lessons are:
\begin{enumerate}
\item Credentials are perishable. Predictive power concentrates in the first year, so any lifetime-stock display (Kaggle's tiers, and by the same logic the badges of professional profiles on service-provider platforms) overstates stale signals by construction. The overstatement is not a corner case: in a quarter of AI-era participations by tiered entrants, every medal behind the tier is more than a year old (26\%; \SIname~A5). Our comparison of the official tier with the recency-weighted index puts a number on what that overstatement costs: up to a sixth (13--16\%) of the available information in sample. A recent-performance summary uses no information platforms lack: an index fit before the AI era already outperforms the official tier on AI-era outcomes. Nor is it what the platform already computes: its own decaying points ranking, reconstructed as of each competition's start, explains less AI-era variation than the frozen index and no more than the lifetime tier. The screening exercise bounds what that information recovery buys: a better broad screen, not a better instrument for the extreme top performers, where lifetime honors still win. Switching the screen from the tier to the index raises the selected entrants' mean performance by 1.3 and 2.2 percentile points at the 10\% and 25\% cutoffs, respectively. The implication is a recent-performance summary displayed alongside lifetime honors, not a replacement of them. We do not observe whether buyers would use the recovered information, and the comparison is a forecasting exercise under the current rules, not a prediction of how participants would behave if the display changed.
\item Credential stocks are stranded by institutional exit: when a platform winds down an evaluation format, its credential stock ages out under the same decay pattern with only a moderate change in the medals' own informativeness, and the credentials stay on display at face value. We do not claim the surviving format produces the better signal: the two formats were statistically indistinguishable in informativeness before the AI transition (0.067 against 0.065), and the platform's shift to execution-based evaluation was already underway years before generative AI arrived. A platform that retires an evaluation format faces a stranding problem that the records themselves do not reveal.
\item Signals live in portfolios: when one signal stopped being issued, statistical weight shifted to the others, and any audit that reads credentials one at a time instead of the full profile will misattribute such shifts to the credentials themselves.
\end{enumerate}

Several limitations bound these lessons. All estimates here are associational: they measure how well credentials predict performance, not why. We also do not observe the demand side: no buyer or employer is seen responding to a credential, so the paper establishes what the credentials predict rather than how they are used, and that demand response is the natural next step. One further gap is an extension rather than a flaw: Kaggle is one platform, chosen because it ran two evaluation formats side by side and phased one out, which is what makes the audit possible. The three lessons are properties of how credentials are aggregated and displayed, so they should carry to other platforms with publicly displayed lifetime credentials that buyers use to screen providers. The finding that verified signals kept their informativeness through the AI transition is narrower. Cheap talk, such as a polished résumé, loses its value as a signal once AI can produce the polish for anyone. A hard-earned credential is scored against withheld answers, which polish cannot move. The same analyses can be repeated on any such platform if the data become available. Two limitations concern the AI-era sample itself. First, the AI-era sample is shaped by survivorship: most pre-AI participants did not compete again, and those who did were disproportionately the credentialed (section~\ref{sec:revaluation}), so most AI-era estimates describe the minority who kept competing. Among users who competed in both eras, the decline of the upload-earned stock is a third of its full-sample size but still significant; selection can be bounded, not eliminated (\SIname~A4). Second, the AI-era window is short: 2023--2025 supplies three years of outcomes for a decay process whose timescale is one year.

\subsection*{Acknowledgments}

AI-assistance disclosure: analyses and drafting used AI coding assistants; all results were verified against the underlying data; any remaining errors are ours.

\bigskip\bigskip

\begin{center}
{\LARGE Appendix}
\end{center}
\renewcommand{\thesection}{A\arabic{section}}
\setcounter{section}{0}
\renewcommand{\thetable}{A\arabic{table}}
\setcounter{table}{0}
\renewcommand{\thefigure}{A\arabic{figure}}
\setcounter{figure}{0}

\section{Pre-committed predictions and their outcomes}
\label{si:predictions}

All predictions were committed to the project's git repository before the corresponding estimation code existed; the commit ordering in the history is the pre-commitment record. \precommitrecord{} Table~\ref{tab:predictions} records every prediction and its outcome; the paper reports the rejections as rejections. The rejection that most shapes the paper is the age-profile prediction (P5a). We had predicted that the age-decay curves themselves steepened at the AI transition. Instead, about half of the ``credential collapse'' traces to ordinary aging of an unreplenished stock. Five exercises were not pre-committed. The decomposition of old-medal gains (main-text Fig.~4) was added after questioning our own earlier results (the age-profile findings of the P5 rows). The windowed-tier comparison and the screening exercise (section~\ref{si:tiers}) were added to separate what the recency-weighted index recovers from how a platform could use it. The benchmark against the platform's decaying points ranking (section~\ref{si:tiers}) was added to test the index against the recent-performance summary the platform already computes. The stale-display count (section~\ref{si:tiers}) was added to state how often a displayed tier rests on medals older than a year. All five are labeled post hoc where they appear, and their post hoc status is recorded in the analysis metadata and the repository history.

\begin{table}[h]
\centering\small
\begin{tabular}{p{9.0cm}p{5.8cm}}
\toprule
Prediction (committed before estimation) & Outcome \\
\midrule
P1a: credential informativeness declines in the AI era; the medal-count slope is smaller for AI-era outcomes & Supported for the upload-earned stock; reversed for the code-earned stock \\
P1b: the decay concentrates in upload-earned credentials; code-earned credentials retain their signal & Rejected when medals are compared at fixed age \\
P1c: decay strongest when predicting code-competition outcomes & Not supported \\
P2a--c: lower returns to an AI-like working style in code than in upload formats, within person & Rejected (precise null) \\
P3a--b: failure rates rise across entry cohorts and with an AI-like working style & Rejected: the raw rise reflects which competitions are entered; within a competition the gradient is null or slightly reversed \\
P3c: the AI-era growth in bad outcomes comes from runs that crash and receive no score, rather than from scored runs that drop sharply from the provisional to the final standings & Pattern present in raw shares only \\
P5a: credential age-decay curves steepened at the AI transition & Rejected (no curve-wide steepening) \\
P5b: the age-decay result survives the balanced panel of both-era users & Supported \\
\bottomrule
\end{tabular}
\caption{Pre-committed predictions and outcomes. \predictionsdoc{} Identifiers are the project's internal labels (P1--P3 for the three core analyses, P5 for the staleness-test addendum; no P4 set was defined). ``Rejected'' denotes statistically precise evidence against the stated direction; ``Not supported'' denotes absence of the predicted pattern.}
\label{tab:predictions}
\end{table}

\section{Data and sample construction}
\label{si:data}

\textbf{Source.} Meta Kaggle, the platform's public archive (snapshot of July 29, 2026): 18,668,885 competition submissions, 9.3 million teams, 9.5 million team memberships, and 11,930 competitions, 2010--2026. \textbf{Competition taxonomy.} The medal-eligible (``tier-eligible'') universe comprises 460 competitions. Evaluation formats are classified as \emph{simulation} (the archive records their matches as ``episodes''), \emph{code} (only notebook submissions, the platform's ``kernels'', are accepted), or \emph{upload} (the remainder); the classification was checked against a fixed list of twelve widely known competitions spanning all three formats (12/12 correct; the list is in the archived taxonomy script). Two-stage and forecasting competitions (whose final scores depend on data generated after the deadline, so movement between the provisional and final standings is by design) were identified by a rules-text screen plus hand curation of all 27 candidates (19 flagged), each with quoted rules evidence in the curation script; they are excluded throughout. Synchronous code execution (the platform runs the code at submission) launched on May 17, 2019 (the launch competition's platform record matches the announcement date); scoreless in-deadline submissions are 0\% of code-competition entries before that date, 11--22\% after, and $\approx$0\% in upload-competitions throughout, which validates scoreless entries as the failed-run measure. \textbf{Points ranking.} Alongside tiers, the platform ranks members by competition points that decay: a competition awards $[100000/\sqrt{N_{\text{teammates}}}]\,\text{Rank}^{-0.75}\,\log_{10}(1+\log_{10} N_{\text{teams}})\,e^{-t/500}$ points, with $t$ the days since its deadline, a half-life of about 346 days (\texttt{kaggle.com/progression}). The formula was revised in May 2015 and predates the tier system introduced in July 2016. A member's profile page lists the tier, the current and highest rank, and the medal counts together, while the tier alone marks the member's username across the site. Every tier-eligible competition in the archive carries a positive ranking multiplier (14 carry a multiplier other than one, the latest in 2019); the published ranking lists members at Expert or above, while the points accrue to everyone. We reconstruct each member's points as of each competition's start from the archive's private leaderboard ranks, team memberships, ranked-team counts, deadlines, and multipliers, counting every prior ranked finish in a tier-eligible competition with a deadline strictly before the start (section~\ref{si:tiers}). Validated against members' current points, the floor of the reconstructed total matches the archive's integer field to within 5\% for 93\% of the 210,249 members with positive points (49\% exactly; Spearman 0.995), once the field's timing is taken into account: it runs 11 days behind the file date, an offset calibrated by centering the reconstruction's error, which rescales every member's points by one factor and cannot change any ranking. Members whose ranked finishes lie only in the 29 competitions that award ranking points but no medals (809 members, median 6 points) count as reconstructed zero. \textbf{Analysis panel.} One row per (user, competition) roster participation: 444,698 participations, 197,561 users, 214 upload- and code-competitions (95 upload, 119 code) with ranked non-benchmark teams (benchmark entries are leaderboard baselines, not competitors), deadlines 2018--2025, the 2022Q4 transition quarter excluded. Outcomes are team results attributed to roster members (solo-only and submitter-attributed variants in section~\ref{si:robust}). Credential counts are reconstructed exactly as they stood at each competition's start, built from medal award dates, counted once per team, and split by format of earning and age band; the age bands partition each total by construction (enforced by an assertion in the construction code). \textbf{Entry cohorts and failed runs.} Entry cohorts group users by the date of their first in-deadline submission: 2015--2018 (the pre-AI cohort), 2019 to June 28, 2021 (the transition cohort), and June 29, 2021 onward (the AI-era cohort), the release date of GitHub Copilot; earlier entrants are excluded. Cohorts split at Copilot because code assistance reached entrants before chat assistance, while outcome eras split at ChatGPT, the shock to the general entrant population. Failed runs are scoreless in-deadline submissions in synchronous code-competitions, the mode in which the platform runs the code at submission.

\section{The working-style index}
\label{si:signature}

The behavioral index mirrors the design validated in \citet{yao2026scaffold}. Three submission-telemetry components (the percentile of the first scored submission against the competition's final public leaderboard, $+$; median submissions per competition, $-$; same-day resubmission share, $-$) are aggregated per user-quarter, sign-aligned, standardized, and combined by the first principal component, with loadings frozen on a broad reference sample (explained variance 0.46, where three uncorrelated components would each give one third; all loadings positive). Construct validation: the two iteration components correlate with pre-AI-era credentials in the direction \emph{opposite} their AI-like signs, while the first-shot component partially proxies skill; accordingly, the confound-robust variant is built from the two iteration components only, dropping the first-shot component (correlation 0.98 with the full index) and is the primary specification wherever the index appears. Validation scope: these exercises establish the index as a working-style measure; whether it responds to AI adoption is not identified, so a null for the index is a null for working style rather than a direct test of AI use. Within-user event-time paths for the components show no break at the release of ChatGPT that stands out from pre-release fluctuations (post-release coefficients are of the same sign and size as multiple pre-release ones; no formal break test is imposed), and the index is not a measure of copying: how often a user builds on others' shared notebooks (``forking'') correlates with the index components at or below 0.16. Analyses use the trailing four-quarter index, excluding the outcome competition's quarter by construction.

\section{Decomposition, survivorship, selection, bounding}
\label{si:robust}

\textbf{Stock-change decomposition.} The decomposition of each stock's slope change is a band-level Oaxaca--Blinder-style split \citep{oaxaca1973male, blinder1973wage}: composition re-weights band shares at pre-AI slopes, the pure curve shift moves band slopes at the pre-AI mix, the interaction takes their product, and a band-level aggregation residual closes the gap to the total-stock slope. The paper's aging share is the composition component alone, 51\% of the upload-stock decline (bootstrap 95\% interval 42--61\%). The interaction and the residual can be assigned to aging, to the curve shift, or to neither, and the decomposition itself does not say which; assigning them to aging would put the share at 38\% (interaction only; 31--49\%), 85\% (residual only; 78--96\%), or 73\% (both; 64--87\%), so the identified statement is the composition component. A competition-cluster bootstrap (resampling competitions with replacement, $B=200$) refits the stock and band regressions on each draw and yields percentile intervals for the decomposition components and for each share. \textbf{Survivorship.} Continued participation into the AI era is steeply selected on credentials: retention rates by pre-AI-era medal count are 7.8\% (no medals), 24.6\% (1--2), 39.3\% (3--5), and 56.6\% (6+). In the balanced panel of users with participations in both eras (124,327 participations), the decline of the upload-earned stock is $-0.019$ ($p=0.001$), a third of the full-sample $-0.055$, and the rise of the code-earned stock is $+0.014$ ($p=0.13$) against $+0.022$; the panel excludes AI-era newcomers as well as non-continuers, so the two samples differ in composition, not only in selection. The upload decline ranges from $-0.039$ to $-0.071$ under best/worst-case trimming of 5--15\% of AI-era outcomes among medaled users (a sensitivity adaptation of the trimming bounds of \citet{lee2009training}; the formal bound applies to mean differences, not slopes). Unconditional old-upload-medal slopes rise in the AI era (main-text Fig.~4A), but the balanced panel speaks against survivor composition as the driver: re-estimated among both-era users, the badge-count-read gains are unchanged ($+0.061$ and $+0.037$ for the 1--2 and 2--3 year bands, both $p<0.001$), so the rise reflects stale medals proxying for their holders' surrounding profiles; no unconditional rise appears for code-earned medals (main-text Fig.~4B). \textbf{Selection into formats.} Format choice in the AI era sorts primarily on legacy credentials (upload-credentialed users' code share is 11.3 percentage points lower per log medal) and only mildly on the working-style index; the within-person null holds among the 2,431 users active in both formats ($+0.005$, $p=0.40$) and within years. \textbf{Attribution.} The decay results are stable across solo-only participations, submitter-attributed participations, and team-size controls. \textbf{External check.} In community competitions, a disjoint outcome universe that never awards tier medals, the same stock pattern appears for AI-era outcomes (upload-earned slope change $-0.045$, $p=0.002$; code-earned $+0.073$, $p<0.001$; $n=344{,}181$). \textbf{Sensitivity to unobservables.} A coefficient-stability calculation in the style of \citet{oster2019unobservable} on the within-$R^2$ basis yields a breakdown $\delta$ of 0.44 for the upload decay interaction; following \citet{masten2026sign}, we treat explain-away statistics as unable to certify sign robustness and report the exercise as disclosure rather than evidence; the balanced-panel, trimming, attribution, and external checks above carry the robustness argument.

\textbf{Old-medal revaluations by read.} Fig.~4 of the main text shows the AI-era changes in old-medal slopes as controls move from the badge-count read to the full-profile read. As an analogy for the code-earned pattern, imagine two recommendation letters from referees who always agree with each other: the second letter adds little beyond the first. Once the more discerning and hence more informative referee's letter arrives, the other letter matters only where it deviates, and extra praise that the discerning referee does not echo can even count against the candidate. For code-earned medals, the AI-era gains shrink with fewer controls and are absent under the badge-count read ($+0.002$, $p=0.84$, and $-0.016$, $p=0.20$). Medals earned \emph{before} ChatGPT show the same re-weighting ($+0.052$, $p<0.001$, and $+0.026$, $p=0.035$, controlling post-ChatGPT code medals; main-text Fig.~4B, rightmost group), so a changed pool of medal earners cannot explain the gains. \textbf{Negative conditional weights.} For a standardized outcome and two standardized signals with correlations $r_{y1}$, $r_{y2}$, and $r_{12}$, the joint OLS coefficient on the second signal is $\beta_2=(r_{y2}-r_{y1}r_{12})/(1-r_{12}^2)$, which is negative exactly when $r_{y2}<r_{y1}r_{12}$: a signal whose own outcome correlation falls short of the product of its correlation with a stronger signal and that signal's outcome correlation takes a negative conditional weight, the configuration known as suppression \citep{conger1974suppressor}. With additional controls the same condition applies after partialling them out. The old-code pattern in the main text fits this configuration: unconditional slopes of $+0.033$ that turn negative only once the upload stock and prior participation enter.

\section{Tier reconstruction, the tier-versus-index comparison, and the post hoc exercises}
\label{si:tiers}

\begin{figure}[t]
\centering
\includegraphics[width=\singlefigwidth]{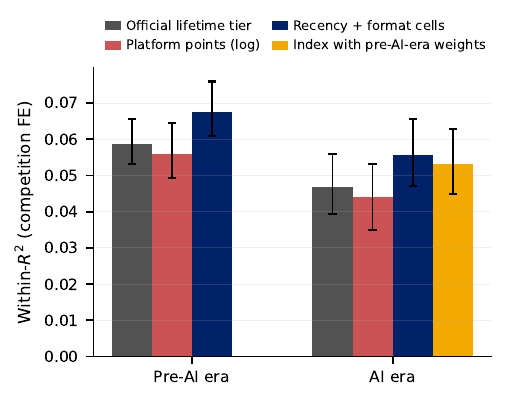}
\caption{The official tier discards available information. Within-competition $R^2$ of hidden-test performance on the official lifetime tier (reconstructed as of each competition's start) versus the recency-weighted index of the same medal events (weights by age band and format), by outcome era; the gold bar reports the performance of an index whose weights were fit on pre-AI-era outcomes only and evaluated on AI-era outcomes. Whiskers are competition-cluster bootstrap 95\% intervals ($B=200$). Paired intervals exclude zero for three differences: the cells minus the tier in both eras ($0.007$--$0.011$ pre-AI, $0.006$--$0.012$ AI era), the pre-AI-weighted index minus the tier on AI-era outcomes (the gold bar against the AI-era tier bar, $0.004$--$0.009$), and that index minus the points ($0.004$--$0.013$). The interval for the points minus the tier includes zero in each era ($-0.007$ to $0.001$ in both).}
\label{fig:aggregator}
\end{figure}

Kaggle's competition tiers are deterministic in medal counts. Encoding both candidate rule variants and validating against observed current tiers selects the cascade rule (higher medals satisfy lower requirements): Expert requires two medals, Master one gold and two silver-or-better, Grandmaster five gold including one solo gold. Validation against observed current tiers uses the full confusion matrix, not recall alone: recall is 94.9\% (Expert), 98.0\% (Master), and 98.5\% (Grandmaster); the false-positive rate among 31,316 confirmed untiered users is 1.3\%; overall accuracy across all four classes is 97.7\%, and the rule variant is selected on accuracy. Every medal event in the archive carries an award date, so the tier as of any date is computable for every participation without timing ambiguity. The remaining mismatches are unexplained. The joint specification regresses hidden-test performance on (i) dummies for the official tier as of the competition's start, (ii) medal counts by age band and format, and (iii) both, with competition fixed effects, separately by era; adding the tier to the cells raises within-$R^2$ by 0.004 (pre-AI era) and 0.003 (AI era), while adding the cells to the tier raises it by 0.013 and 0.012. The era-generalization index fixes the cell weights on pre-AI-era outcomes only and is evaluated out of era on AI-era outcomes (Fig.~\ref{fig:aggregator}). \textbf{Inference and addenda.} Uncertainty for every within-$R^2$ comparison is a cluster bootstrap resampling competitions with replacement ($B=200$, seed recorded in the analysis metadata; each drawn competition receives a distinct fixed-effect identifier; the pre-fit index weights are held fixed, so the out-of-era interval reflects evaluation-sample uncertainty for this already-fixed formula, not first-stage weight-estimation uncertainty; percentile bounds at this replication count are approximate). The 95\% intervals: cells minus tier, $0.007$--$0.011$ (pre-AI era) and $0.006$--$0.012$ (AI era); the pre-AI-weighted index minus the tier on AI-era outcomes, $0.004$--$0.009$; the share of the cells' content the tier discards, 10.8--15.8\% and 11.4--19.4\%. A fresh-band ablation gives the concentration claim its information form: the two under-one-year bands alone deliver 99.4\% (pre-AI era) and 95.3\% (AI era) of the within-$R^2$ of all nine cells (bootstrap intervals 98.8--99.6\% and 93.9--96.3\%).

\textbf{Windowed tiers and the screening exercise (post hoc).} Neither exercise was pre-committed (section~\ref{si:predictions}). \emph{Windowed tiers.} The cascade rule is recomputed counting only medals awarded within the trailing 24 or 12 months before each competition's start, with the lifetime thresholds unchanged, so each result is one particular rolling rule rather than a general test of recency. Windowing empties the upper classes: in the AI era the 12-month rule leaves 64 participations at Grandmaster and 4,056 at Master, against 1,948 and 7,126 under the lifetime rule. Within-competition $R^2$ on AI-era outcomes is 0.0459 (24 months) and 0.0392 (12 months) against 0.0469 (lifetime); paired bootstrap 95\% intervals for the differences from the lifetime rule are $-0.0027$ to $0.0003$ and $-0.0105$ to $-0.0054$, and the frozen index exceeds the 24-month and 12-month rules by 0.0053--0.0098 and 0.0115--0.0169. Dropping the solo-gold requirement for Grandmaster changes these values by less than 0.0001. \emph{Screening exercise.} Within each AI-era competition, entrants are ranked by each selector (lifetime tier, 12-month tier, and the raw frozen index) and the top $\lceil kn/100\rceil$ are selected for $k=10$ and $k=25$, so set sizes are identical across selectors. Ties at the cutoff, which occur in every competition for the coarse tier at $k=10$, are integrated exactly: entrants tied at the cutoff share the remaining slots with equal fractional weight, the expectation under uniform random tie-breaking; a single seeded draw with common random priorities differs by at most 0.0007. Outcomes are the selected set's mean performance and its share of top-decile finishers (private percentile at or below 0.10, a team outcome attributed to members), averaged across competitions, against the corresponding values among all eligible entrants (mean performance 0.528; top-decile share 0.139). Index minus lifetime tier: mean performance $+0.013$ (0.006--0.019) at 10\% and $+0.022$ (0.018--0.025) at 25\%; top-decile share $-0.022$ ($-0.033$ to $-0.011$) at 10\% and $+0.010$ (0.007--0.013) at 25\%. A solo-entrant-only variant preserves the ordering for mean performance at both cutoffs and for the top-decile share at 25\%; at 10\% the 12-month tier and the index are within 0.003 of each other in both variants. Intervals come from the same competition-cluster draws as the other comparisons, with per-competition screening records resampled, and the pairing is verified on every draw. The exercise ranks entrants who chose to compete, with team outcomes attributed to members; it cannot establish hiring benefits.

\textbf{Points ranking (post hoc).} Not pre-committed (section~\ref{si:predictions}). The platform's decaying competition points, reconstructed as of each competition's start (section~\ref{si:data}), enter the horse race as a single regressor, $\log(1+\text{points})$, on the same rows and with the same fixed effects and clustering. They explain less than the lifetime tier in both eras (within-$R^2$ 0.056 against 0.059 pre-AI; 0.044 against 0.047 in the AI era; the bootstrap intervals for the difference include zero in both), and raw points explain less still (0.033 and 0.035). On AI-era outcomes the frozen index explains 0.053 against the 0.044 of log points; the paired difference is $+0.009$ (95\% interval 0.004--0.013), and the nine cells exceed log points by 0.012 (0.007--0.016). The exercise was not pre-committed; its decision rule, including a practical-equivalence margin of 0.002 in within-$R^2$, was fixed in the project plan before any predictive number from it was computed, and the plan is part of the archived record. The interval lies outside the margin, so the index is judged to recover information the platform's own summary does not. Descriptive increments, without mechanism claims: adding log points to the tier raises within-$R^2$ to 0.070 (pre-AI) and 0.056 (AI era); adding them to the nine cells, to 0.077 and 0.062. The AI-era points $R^2$ is unchanged under the archive's team-count denominator (0.044), with all multipliers set to one (0.044), with only post-May-2015 finishes counted (0.044), and with a seven-day result-availability lag (0.043). In the screening exercise points enter as a third selector (their ordering is invariant to the log): at 10\% they select entrants with mean performance 0.684 and a top-decile share of 0.262 (index 0.699 and 0.295; tier 0.686 and 0.318), at 25\% 0.641 and 0.212 (index 0.645 and 0.228; tier 0.624 and 0.219); solo-only, 0.652 and 0.175 at 10\%. Paired intervals from the same replayed draws: index minus points, mean performance $+0.015$ (0.011--0.019) at 10\% and $+0.005$ (0.002--0.008) at 25\%; top-decile share $+0.034$ (0.027--0.041) and $+0.017$ (0.013--0.021); points minus tier, top-decile share $-0.056$ ($-0.067$ to $-0.045$) at 10\% and $-0.007$ ($-0.012$ to $-0.002$) at 25\%, mean performance $-0.002$ ($-0.010$ to $0.006$) at 10\% and $+0.017$ (0.011--0.022) at 25\%. The bootstrap replays the competition-cluster draws of the main comparison, verifies the pairing on every draw, and resamples the screening records for both subsets.

\textbf{Stale display (post hoc).} Not pre-committed. A medal is fresh if it is under one year old at the competition's start. Among participations by members holding a tier (Expert or above under the reconstructed competitions-track tier), the share with no fresh medal was 17\% before the AI transition (7,381 of 44,533) and 26\% in the AI era (7,608 of 29,633); among participations by members with any medal, 20\% and 28\%. For these entrants the tier and the medal counts on display rest entirely on medals older than a year.

\section{Person-level nulls}
\label{si:nulls}

\begin{figure}[t]
\centering
\includegraphics[width=\singlefigwidth]{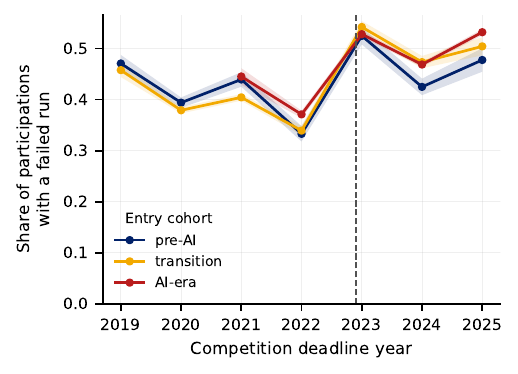}
\caption{Failed-run shares track competitions, not entry cohorts. Share of code-competition participations with at least one failed run, by entry cohort and competition year; cohorts group entrants by first-submission date (pre-AI cohort 2015--2018; transition 2019 to the June 2021 release of Copilot; AI-era after it); shaded bands are 95\% confidence intervals across participations. Cohort curves move together, and the dashed line marks the release of ChatGPT; competition fixed effects alone erase the raw difference of the AI-era cohort relative to the pre-AI cohort ($+0.056$, $p<0.001$, to $-0.003$, $p=0.77$).}
\label{fig:failures}
\end{figure}

Two mechanisms we had predicted would accompany the AI era are absent at the person level. Both predictions had been committed to the version-controlled record before estimation (section~\ref{si:predictions}). First, contrary to our prediction, we find that an AI-like working style predicts performance no differently across the two evaluation formats. We measure this working style with a behavioral index built from submission telemetry. Its three standardized components are submissions per competition, the share of same-day resubmissions, and the strength of a user's first scored submission against that competition's final public leaderboard. The first two enter negatively, so a high index means few iterations. The third enters positively, so a high index also means a strong first shot. Both patterns are expected when AI drafts much of the work. The primary version of the index drops the first-shot component, which partially proxies skill, and keeps the two iteration components. The resulting index correlates at 0.98 with the full three-component version, so the drop changes the index little. As a construct check, the iteration components correlate with pre-AI-era credentials in the direction opposite their AI-like signs: credentialed users iterate more, not less (section~\ref{si:signature}). Under person and competition fixed effects, the index's return in code-competitions relative to upload-competitions is statistically indistinguishable from zero ($+0.006$ per index unit, s.e.\ 0.005, $p=0.28$). The null is precisely estimated: even at the edge of the 95\% confidence interval, two entrants one standard deviation apart in working style (sample standard deviation 1.2) would differ in their code-versus-upload performance gap by at most about $0.02$, roughly an eighth of the main text's fresh-medal slope (0.15--0.18). The null persists among the 2,431 users who compete in both formats, for whom the cross-format comparison is purely within person (section~\ref{si:robust}). The index captures a style of working; whether an entrant actually used AI is never observed. The null therefore covers the style, and it does not test whether the return to AI use itself differs across the two formats. Second, we predicted that failed runs (code-competition submissions that crash or time out during official execution, observable as scoreless entries) would rise sharply across entry cohorts, and did find the pattern in raw terms. However, Fig.~\ref{fig:failures} shows why: the failure rates of successive entry cohorts (pre-AI, transition, and AI-era cohorts; entrants grouped by the date of their first submission) move together year by year, and the raw difference of AI-era entrants relative to the pre-AI cohort ($+0.056$, $p<0.001$) is erased by competition fixed effects alone ($-0.003$, $p=0.77$), before any individual controls enter. The transition cohort even shows a small decline rather than a rise ($-0.014$, $p=0.04$). In short, failures track what competitions have become rather than who the entrants are; conditional on the competition, no cohort fails more than its predecessors.

\section{Code, reproducibility, and AI assistance}
\label{si:repro}

All data are public (Meta Kaggle). \repostatement{} Each stage of the analysis underwent an independent adversarial code review; review reports and their dispositions are part of the repository record. Analyses and drafting used AI coding assistants; every reported number traces to an archived results file generated by the reviewed pipeline.

\singlespacing
\bibliographystyle{chicago}
\bibliography{Bibliography_base}

\end{document}